# DASentimental: Detecting depression, anxiety and stress in texts via emotional recall, cognitive networks and machine learning

A. Fatima[1], Y. Li[2], T.T. Hills[3] and M. Stella[1]

1 CogNosco Lab, Department of Computer Science, University of Exeter, UK
2 Max Planck Institute for Human Development, Germany
3 Department of Psychology, University of Warwick, UK
* Corresponding Author: m.stella@exeter.ac.uk.

October 26, 2021

**Abstract**

Most current affect scales and sentiment analysis on written text focus on quantifying valence (sentiment) – the most primary dimension of emotion. However, emotions are broader and more complex than valence. Distinguishing negative emotions of similar valence could be important in contexts such as mental health. This project proposes a semi-supervised machine learning model (DASentimental) to extract depression, anxiety and stress from written text. First, we trained the model to spot how sequences of recalled emotion words by N=200 individuals correlated with their responses to the Depression Anxiety Stress Scale (DASS-21). Within the framework of cognitive network science, we model every list of recalled emotions as a walk over a networked mental representation of semantic memory, with emotions connected according to free associations in people's memory. Among several tested machine learning approaches, we find that a multilayer perceptron neural network trained on word sequences and semantic network distances can achieve state-of-art, cross-validated predictions for depression ($R = 0.7$), anxiety ($R = 0.44$) and stress ($R = 0.52$). Though limited by sample size, this first-of-its-kind approach enables quantitative explorations of key semantic dimensions behind DAS levels. We find that semantic distances between recalled emotions and the dyad "sad-happy" are crucial features for estimating depression levels but are less important for anxiety and stress. We also find that semantic distance of recalls from "fear" can boost the prediction of anxiety but it becomes redundant when the "sad-happy" dyad is considered. Adopting DASentimental as a semi-supervised learning tool to estimate DAS in text, we apply it to a dataset of 142 suicide notes. We conclude by discussing key directions for future research enabled by artificial intelligence detecting stress, anxiety and depression.

# Introduction

Depression, anxiety and stress are three negative emotions that are associated with different experiences and psychopathological consequences [1, 2, 3]. According to the Depression, Anxiety and Stress Scale (DASS) [4], depression is associated with profound dissatisfaction, hopelessness, abnormal evaluation of life, self-deprecation, lack of interest, and inertia. Similarly, anxiety is associated with hyperventilation, palpitation, nausea and physical trembling, and stress is associated with difficulty relaxing, getting easily agitated, and impatience (cf. [1, 5, 6]).

While these emotions are discrete and highly complex [3, 6], they vary along a primary and culturallyuniversal dimension of valence, i.e. perceived pleasantness [2, 7, 8]. The universality of valence is behind its frequent use in self-report affect scales [9]. In particular, valence (i) can be easily quantified along a continuous scale and (ii) explains the largest variance when compared to other proposed emotional dimensions, including arousal [7]. Unfortunately, valence is often the only output of affect scales. This is potentially problematic for measuring depression, anxiety, and stress, which are more complex [10]: these three distinct types of psychological distress are similar in valence but widely different along other dimensions like arousal [11]. In fact, depression, anxiety, and stress are difficult to distinguish using valence alone [12, 13, 3]. This underlines the need for richer mappings between emotional dimensions and depression, anxiety and stress (DAS) levels.

In the present study, we examine a new approach for detecting DAS levels by exploring how they are related to the emotions people recall when asked to report how they felt recently [8]. In particular, we develop a machine learning approach (DASentimental) for extracting more comprehensive information of reported emotions from a recall-based affect scale, i.e. the Emotional Recall Task [8]. By using DASentimental to extract information from recently experienced emotions, we show how DASentimental can be used to make inferences about DAS levels that extend beyond valence and, subsequently, investigate natural language more generally. In what follows, we briefly review the literature on making inferences from affect scales and then describe our new approach based on machine learning.

**Literature review: Cognitive data science, mental well-being and issues of affect scales**

Mental well-being is a psychological state in which individuals are able to cope with negative stimuli and emotional states [14, 15, 3, 6]. Assessing the connection between emotions and mental well-being is a key yet relatively unexplored dimension in *cognitive data science*, i.e. the branch of cognitive science investigating human psychology and mental processes under the lens of quantitative data models [16, 8, 17, 18, 19, 20, 21, 22, 23, 6].

Anxiety, stress and depression can impair mental well-being, with consequences as extreme as ending one's own life [5, 24]. The early quantitative recognition of distress signals that might affect mental health is crucial to providing support and boosting wellness. Recognition starts with diving deep into the mindset of individuals and understanding the emotions they are going through. Psychological research [8, 22, 15, 25, 26] has recognised that our mental state relates to the way we humans communicate, hence our written and spoken language can reveal our psychological states. Thus, the emotional state of an individual can be anticipated through their communication [22, 27]. Identifying a quantitative coexistence and correlation among emotional words used by an individual can unveil crucial insights about their emotional state [28, 29]. However, using only valence, also called sentiment in Computer Science [30], to assess DAS levels is likely to be insufficient. Capturing how people reveal various forms of emotional distress in their natural language is therefore an important and open research area.

Outputs of affect scales typically include scores that quantify emotional valence. For example, the Positive Affect and Negative Affect Scale (PANAS; [13]), arguably the most popular self-report affect scale, asks people to evaluate their emotional experience against a predetermined emotion checklist that contains 10 positive words and 10 negative words (e.g. to what extent did you feel irritated over the past month?). By summing up the responses, the PANAS provides two scores, one for positive affect (PA) and one for negative affect (NA). The PANAS essentially splits the emotions in two groups based on valence, and consequently ignores the within-group difference in valence. That means, for example, emotions in the negative affect list such as *guilty* and *scared* are treated as if they have the same emotional impact.

Understanding mental well-being could be enhanced both by investigating richer sets of emotions— including everything a person might remember about their recent emotional experience—and by examining the sequence of those emotional states. A precedent for this approach was recently set by the publication of the Emotional Recall Task (ERT) [8]. The ERT asks participants to produce 10 emotions that described their feelings. The sequences of words produced in the ERT represents a potential wealth of information for adapting machine learning to sentiment analysis. The idiosyncratic features of these individual words may contain information beyond valence. Indeed, arousal is often included as a primary predictor in addition to valence, for example, in the two-dimensional circumplex model of emotions [11]. Yet the ERT is likely to contain other dimensions as well. For example, anger and fear are both highly negative and highly arousing, but they refer to different experiences and prepare people for different sets of behaviors with, for example, anger triggering potential aggression while fear triggering either freezing or fleeing. The order in which words are recalled in the ERT may also contain useful information, as it indicates the availability of different emotions and may therefore signal information about emotional importance [31, 32]. For example, earlier-recalled words are likely to provide more information to well-being than later-recalled emotions. Finally, the ERT may also contain information on emotional granularity, i.e. a psychological construct referring to the individuals' ability to discriminate between different emotions [14]. For example, a person with high (as opposed to low) emotion granularity would tend to use more distinct words, like 'anxious' (as opposed to 'bad'). It was found that people with higher emotional granularity reported better well-being, less prone to mental illness, probably because the sophisticated understanding of one's negative emotions create better coping strategies[14]. Crucially, people with less emotional granularity were found to be more likely to focus on valence and use happy and sad to cover the entire spectrum of positive emotions and negative emotions [2]. All these patterns and strategies represent the building blocks of our approach with DASentimental.

## Research aims

This project focuses on sequences of emotional words, whose ordering and semantic meaning contain features that are assumed to be predictive of stress, anxiety and depression. Having defined these psychological constructs along the psychometric scale represented by DASS, the current project aims to reconstruct the model between emotional word sequences and DAS levels through machine learning. We adopt a semi-supervised learning approach mainly composed of two stages. Firstly, we train a machine learning regression model over cognitive data coming from the ERT task [8]. Through cross-validation and feature selection, we enrich word sequences with a cognitive network representation [15, 33] of semantic memory. We show that semantic prominence in the recall task as captured by network degree can boost the performance of the regression task. Having selected the best performed model, we apply it to identifying emotional sequences in text, providing estimations for the DAS levels of narrative/emotional corpora like, for instance, suicide notes [28, 24].

We conclude our investigation with a discussion about the cognitive relevance of models tested here, and the limitations and future research directions opened by our approach.

## Methods

This Section outlines the datasets and methodological approaches adopted in this manuscript.

### Datasets: Emotional recall data, free associations and suicide notes

Four datasets were used to train and test DASentimental: (i) the Emotional Recall Task (EAT) dataset [8], (ii) the Small World of Words free association data for English [33], (iii) the corpus of genuine suicide notes curated by Schoene and Dethlefs [28] and (iv) valence-arousal norms by Mohammad [10].

The ERT dataset is a collection of emotional recalls provided by 200 individuals and matched against various psychometric scales like the DASS (Depression, Anxiety and Stress Scale) one [4]. During the recall task, each participant was asked to produce a list of 10 words expressing the emotions they felt in the last month. Participants were also asked to assess items on psychometric scales, thus providing data in the form of word lists/recalls, e.g. {*anger, hope, sadness, disgust, boredom, elation, relief, stress, anxiety, happiness*}, and psychometric scores, e.g. anxiety/depression/stress levels between 0 (low) and 20 (high). The dataset is completely anonymous and it enables the creation of a mapping between the sequences of emotional words recalled by individuals and their mental well-being, which is built here through a machine learning approach.

The Small World of Words [33] is an international research project aimed at mapping human semantic memory [19] through free associations, i.e. conceptual associations where one word elicits the recall of other ones [33]. Cognitive networks made of free associations between concepts have been successfully used to predict a wide variety of cognitive phenomena, including language processing tasks [33], creativity levels [18, 20], early word learning [34, 35], picture naming in clinical populations [36] and semantic relatedness tasks [37, 38]. Being free from specific syntactic or semantic constraints, free associations capture a wide variety of associations encoded in the human mind [39] and this element, together with the many successful applications listed above, motivated our choice of using free associations for modelling the structure of semantic memory upon which the ERT recalls are selected from. This modelling approach posits that: (i) all individuals, independently on their well-being, possess a common structure of conceptual associations, and that (ii) the connectivity of emotional words is not uniform but there are more (and less) well connected concepts. Although preliminary evidence shows that semantic memory might be influenced by external factors like distress [40] or personality traits [41], we have to adopt point one as a necessary modelling simplification in absence of free association norms across clinical populations. The adoption of a network structure and the second point operationalise the task of identifying how semantically related emotional words are in terms of network distance, i.e. the length of the shortest path connecting any two nodes [35]. Network distance in free associations was found to outmatch semantic latent analysis when modelling semantic relatedness norms [38, 37], supporting our approach.

The corpus of suicide notes is a collection of 142 suicide notes by people who ended their lives [28]. The dataset was curated by and analysed for the first time by Schoene and Dethlefs [28], who used it to devise a supervised learning approach to automatic detection of suicide ideation. These notes were collected from various sources, including newspaper articles and other existing corpora. All the notes are anonymised by removing any links to a person or place or any other identifying information. Already investigated in previous studies under the lens of sentiment analysis [28], cognitive network science [24], and recurrent neural networks [29], this dataset represents here a clinical case study where DASentimental will be applied, once trained on word sequences from the ERT data.

The valence-arousal norms used here indicate how pleasant/unpleasant (valence) and how exciting/inhibiting (arousal) words are when identified in isolation within a psychology mega-study [10]. This dataset included valence and arousal norms for over 20,007 English words and it was used for validating, through the circumplex model of affect [11], results based on DASentimental and text analysis.

**Machine learning regression analysis**

Our DASentimental approach aims at performing the extraction of anxiety, depression and stress levels from a given text, through semi-supervised learning. DASentimental is a regression model, trained on features extracted from emotional recalls (EAT data) and obtained from the a network representation of semantic memory (free association data). The model is validated against psychometric scores from the DAS scale [8]. By using cross-validation and feature importance analysis [42], we select a best-performing model to detect depression, anxiety and stress levels in previously unseen sequences of words, i.e. texts. All in all, the pipeline implemented in this project can be divided in 4 main sub-tasks, performed one after the other:

1. Data cleaning and vectorial representation of regressor (features) and response (DAS levels) variables;
2. Training, cross-validation and selection of the best performing regression model for estimating DAS levels from ERT data;
3. Estimating the DAS levels of suicide notes by parsing the sequences of emotional words mentioned in each letter;
4. Validating the labelling predicted by DASentimental through independent affective norms [10].

**Data cleaning and vectorial representation of regressor variables**

Our regression task is relative to building a mapping between depression (anxiety, stress) scores $\{Y\}_i$ and features extracted from sequences of emotional words, $\{X\}_{\{i,j\}}$. Each sequence contains exactly 10 words that was produced by a respondent responded in the ERT, e.g. $X_i=\{$*anger, hope, sadness, disgust, boredom, elation, relief, stress, anxiety, happiness*$\}$ and thus $X_{ij}=$*anger*, etc..

Analogously to other approaches in natural language processing [20, 42, 23], we adopt a vectorial representation, transforming the 10-dimensional vectors $X_i$ into $N-$ dimensional vectors $B_i$ where the first $K < N$ entries $B_{ik}$ (for $1 \leq k \leq K$) count the occurrence (1,2,3,...) or absence (0) of a word in the original recall list $X_i$. Furthermore, the remaining entries $B_{ik}$ (for $K + 1 \leq k \leq N$) are relative to additional features extracted from recall lists when embedded in the cognitive network of free associations.

The representation of word lists as binary vectors of word occurrences is also known as Bag-Of-Words (BOW) [43] and it is one of the simplest and most commonly used numerical representations of texts in natural language processing. The representation of word lists as features extracted from a network structure is also known as network embedding and, in cognitive network science, it has been used for predicting creativity levels from animal category tasks [20]. We will provide more info about which network metrics were adopted in this work in the following subsection.

BOW representations can be quite noisy because of different word forms indicating the same lexical item and thus the same semantic/emotional content of a list, e.g. *depressed* and *depression*. Noise in textual data can be reduced by *regularising* the text, i.e recasting different words to the same lemma or form. We cast different forms to their noun counterparts through the WordNet lemmatisation function implemented in the Natural Language Toolkit (NLTK) and available in Python. This data cleaning reduced the overal set of unique words from 526 to $K = 355$ nouns, thus reducing the dimensionality and sparsity of our vector representations.

**Embedding recall data in cognitive networks of free associations**

A crucial limitation of the above BOW representation is that emotional words appearing in any position of a recall/list will have the same weight for the regression analysis. This is in contrast with a rich literature about recall from semantic memory [19, 21], which indicates that in producing list of items from a given category, the first recalled elements are in general more semantically relevant for the category itself. These findings indicate that a better refinement would be to weight word entries in BOW according to their positions in recalls. For instance, the occurrence of "sad" in the first position in recall $i$, i.e. $X_{i1}$ = sad, would receive a higher weight $w_1$ relative to position 1, than if it occurred at later positions. The different weights $\{w\}_j$ could be tailored so that initial words in a recall are more important towards the estimation of DAS levels.

Rather than using arbitrary weightings, we adopt a cognitive network science approach [17]. Emotional words do not come from an unstructured system but are rather the outcome of a search in human memory [21]. Hence, if we model this memory as a network of free associations, then we can embed words in a network structure and measure their relevance in memory through network metrics, cf. [19, 21, 44]. In this way, we can compute the network centrality of all words in a given position $j$ and estimate weight $w_j$ as the average of such scores. This is the approach we adopted in our case.

Our first step was to transform continuous free association data from [33] into a network where nodes represented words and links represented memory recall patterns, e.g. word A reminding at least 2 different individuals of word B. Analogously to other network approaches [38, 37, 35, 34] and because of the asymmetry in gathering cues and targets [39], we considered links as being undirected. This procedure led to a representation of semantic memory as a fully connected network N containing 34,298 concepts and 328,936 links. On this representation we then computed semantic relevance through one local (degree) and one global (closeness) metrics that were adopted in previous cognitive inquiries [36, 18, 34]. Degree captures the numbers of free associations providing access to given concept, whereas closeness centrality identifies how far on average a node is from its connected neighbours (cf. [17]). We checked that all $K = 355$ unique words from ERT data were present in N. Then, we computed degrees and closeness centralities of all words occurring in a given position $j \in \{1,2,...,10\}$ and reported the results in Figure 1.

Although there are several outliers, Figure 1 confirms previous remarks [8] about the ERT data following memory recall patterns with words in the first positions being more semantically prominent than subsequent ones. Since degree and closeness do not seem to display qualitatively different behaviours, we used the median values $m_j$ for degree, see Figure 1 (left), as weights $w_j = m_j$, normalised so that $\sum w_j = 1$. These weights were used to multiply/weight the respective entries of the BOW representation, $B_{ik}$ (for $1 \leq k \leq K$), as to obtain a weighted bag-of-words representation of recalls depending on both the ERT data and the network representation N of semantic memory.

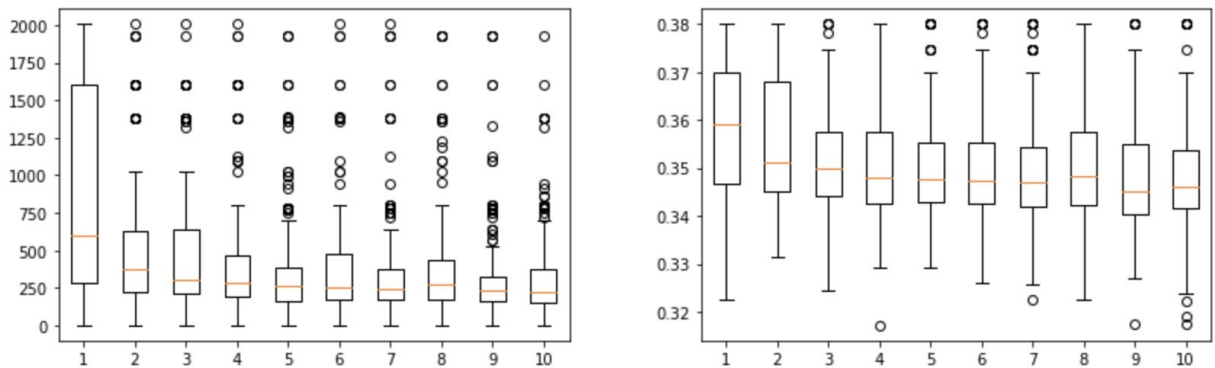

Figure 1: Box plots of degree (left) and closeness centrality (right) of words in the ERT dataset for each position of recall, $j \in \{1,2,...,10\}$.

The above procedure constitutes a first semantic embedding of ERT data in a cognitive network. We also performed a second semantic embedding of ERT recalls by considering them as walks over the structure of semantic memory. Analogously to a previous approach in [20], we considered a list of recalled words as a network path [17], $X_{i1}, X_{i2},...,X_{i10}$, visiting nodes over the network structure and moving along shortest network paths [38]. This second embedding enabled the attribution of a novel set of distance-based features to each recall. In particular, we focused on the following distance-based features:

1. The coverage $C_i$ performed over the whole walk $X_i$ [36], i.e. the total number of free associations traversed when navigating N across a shortest path from node $X_{i1}$ to $X_{i2}$, and then from $X_{i2}$ to $X_{i3}$, and so on. This coverage equals the sum or the total of all the network distances between adjacent words in a given recall.

2. The graph distance entropy [45] $E_i$ of the whole walk $X_i$, computed as the Shannon entropy for the occurrences of paths of any length within $X_i$.

3. The total network distance $D_i$ between all nodes in a walk/recall $X_i$ and the target word "depression". Similarly, we considered also $S_i(A_i)$ as the sum of distances between recalled words and "stress" ("anxiety").

4. The total network distance $H_i$ between all nodes in a walk/recall $X_i$ and the target emotional state "happy". Similarly, we considered also $SS_i(F_i)$ as the sum of distances between recalled words and the target emotion "sad" ("fear").

We considered these metrics based on previous investigations of semantic memory, affect and personality traits. Coverage on cognitive networks was found to be an important metric for predicting creativity levels with recall tasks [16, 20]. A higher coverage and graph distance entropy can indicate sets of responses being more scattered across the structure of semantic memory or oscillating between positive and negative emotional states, with potential repercussions over reduced emotion regulation and increased levels of DAS [8]. Since shortest network distance on free association networks was found to predict semantic similarity [38, 37], we selected semantic distances between recalls and target clinical states as capturing the relatedness of responses to DAS levels. The selection of "happy" and "sad" followed previous results on the circumplex model of affect [11], a model mapping emotional states according to the dimensions of pleasantness and arousal. In the circumplex model, "happy" and "sad" are opposite emotional states and their relatedness to recalls can provide additional information for detecting the presence or absence of states like DAS. We included also "fear" as it is a common symptom of DAS disorders [8].

The validation of these distances as features useful for discriminating different DAS levels is the first point presented in the Results section.

**Model training**

After having obtained weighted and unweighted BOW representations of ERT data, enriched with distancebased measured, we obtained vector representations to be fed to a machine learning regressor.

The following algorithms were tested [46, 42]: (i) decision tree, (ii) multi-layer perceptron, and (iii) recurrent neural network (Long-Short Term Memory or LSTM). Decision tress can predict target values by learning decision rules on how to partition the features of the dataset as to maximise information gain (cf. [23]). The multi-layer perceptron (MLP) is inspired to biological neural networks [47] and consists of multiple computing units, organised in input, hidden and output layers, each one taking a linear combination of features and producing an output according to an activation function. Combinations are fixed according to weights that are updated over time as to minimise the error between the final and target inputs, a procedure that travels backwards on the neural network and known as back-propagation algorithm [46]. LSTM networks feature feed-forward and back-forward loops that interest hidden layers recurrently over training, a procedure known also as deep learning. Additionally, LSTMs feature specific nodes remembering outputs over arbitrary time intervals and this can enhance training by reducing the occurrence of vanishing gradient/getting stuck in local minima.

In the current work, decision trees were trained with *scikit-learn* in Python [42]. For MLPs we selected an architecture using 2 hidden layers with 25 neurons each. A dropout rate of 20% between weight updates in the second hidden layer was fixed as to reduce over-fitting. The number of layers and neurons were fixed after fine-tuning over multiple iterations using the whole dataset of 200 data points and a 4-fold cross-validation. A rectified linear activation function was selected as to keep the output at each layer positive, like DAS scores. For the LSTM architecture, we used 2 hidden layers, each one featuring 4 cells and a dropout rateof 20% to reduce over-fitting.

Training was performed by splitting the dataset of 200 ERT recalls in training (75%) and test sets (25%), according to a 4-fold cross validation. In the regression task of estimating DAS levels from the test set after training, we measured performance in terms of mean squared error (MSE) loss and Pearson's correlation $R$. Vectors of features underwent an L2 regularisation to further reduce the impact of large dimensionality and sparseness during regression. Performance with different sets of featured was recorded as to apply the performing model to text analysis.

**Application of DASentimental to text**

Texts are sequences of words, although in more articulated forms than sequential recalls from semantic memory. Nonetheless, word co-occurrences in texts are not independent from semantic memory structure itself, in fact a growing body of literature in distribution semantics adopts co-occurrences for predicting free association norms themselves [48]. We adopt an analogous approach and use the best performing model from the ERT data to estimate DAS levels in texts based on their sequences of emotional words. DASentimental can thus be considered as a semi-supervised learning approach, trained on psychologically validated recalls and applied to previously unseen sequences of emotional words in texts.

To enhance overlap between the emotional jargon of text and the lexicon of $K = 355$ unique emotional words in the ERT dataset, we implemented a text parser in spaCy, identifying tokens in texts and mapping them to semantically related items in the ERT lexicon. This semantic similarity was obtained as a cosine similarity between pre-trained word2vec embeddings and it is therefore independent from the distances used as features and expressing similarity scores in DASentimental.

As seen in the algorithm reported 1, for every non-stopword [49] of every sentence in a text, the parser identifies nouns, verbs and adjectives and maps them onto their most similar concept, if any, present in the ERT/DASentimental lexicon. This procedure can skip stopwords and enhance the attribution of different word forms and tenses to their corresponding base form from the ERT dataset, which possesses a lesser linguistic variability compared to text because of its recall-from-memory structure.

---

Algorithm 1: Semantic parser identifying emotional words from text that can be mapped onto the emotional lexicon of DASentimental.

---

**Input:** Text from Suicide Note
**Output:** Vector Representation of emotional content, selected words
1 **for each** sentence in suicide note **do** 2 **for each** word in sentence **do** 3 **if** word is negative:
4     isNeg=True
5     **if** word **not** in stopwords **and** word.pos in ['NOUN','ADJ','ADV','VERB']
6     **if** isNeg True:
7     find similar words to the current word antonym in ERT words 8 **if** max similarity$\geq$=0.5:
9         Add most similar word to selected words and update vector
10     isNeg=False
11     **else**:
12     find similar words to the current word in ERT words 13 **if** max similarity$\geq$=0.5:
14         Add most similar word to selected words and update vector
15     **end for** 16 **end for**

---

This mapping is beneficial to making sure that DASentimental does not miss different forms of words or synonyms from texts, ultimately enhancing the quality of the regression analysis. Checking for item similarity in network neighbourhoods drastically reduced computation times, contributing the scalability of DASentimental for volumes of texts larger than the 142 notes used in this first study.

**Handling negations in texts**

SpaCy also provides the ability to track negation in a sentence. Every occurrence of a negation in a sentence was tracked and any occurrence of emotional words in the same sentence was classified, substituting that word to its antonym. A similar approach was adopted in previous studies with cognitive networks [50]. For instance, in the sentence "I am not happy", the word "happy" is not directly checked for similarity against the ERT lexicon. Instead, the antonym of "happy" is found using spaCy, 'sad' in this case, and its similarity is then checked, instead. Handling negations is a key aspect of processing texts. Since more elaborate forms of meaning negations are present in language, this can be considered as a first, simple approach to accounting for semantic negations.

**Psycholinguistic validation of DASentimental for text analysis**

In this first study we used suicide letters as a clinical corpus investigated in previous works [28, 50] and featuring narratives produced by individuals affected by pathological levels of distress. Alas, such corpus did not feature annotations expressing the levels of anxiety, depression and stress felt by the authors of the letter.

Alternatively, we performed emotional profiling [51] over the same set of suicide notes but relying on another psycholinguistic set of affective norms, i.e. the VAD Lexicon by Mohammad [10]. Analogously to the emotional profiling implemented in [30] to extract key states from textual data, we used the VAD Lexicon to provide valence and arousal scores to lemmatised words occurring in suicide notes. We also applied DASentimental to all suicide notes and plotted the resulting distributions of depression (anxiety, stress) scores. A qualitative analysis of the distributions highlighted tipping points, which were used for partitioning the data into letters with high and low levels of estimated depression (anxiety). Tipping points were selected instead of medians because most notes elicited no levels of estimated DAS levels, thus producing imbalanced partitions. These tipping points were identified as being 6 for depression, 2 for anxiety and 4 for stress. As reported in Figure 3 (bottom), above these tipping points the distributions exhibited cut-offs or abrupt changes.

We then compared the median valence and arousal of words occurring in high and low partitions of the suicide corpus. Our exploration was guided by the circumplex model of affect [11], which maps "depression" as a state with negative valence and low arousal, "anxiety" as a state with negative valence and high arousal, and "stress" as a state in-between "anxiety" and "depression". For every partition, our educated guess is for letters tagged as "high" by DASentimental to feature more extreme language.

# Results

This section reports on the main results of the manuscript. Firstly, semantic distances and their relationships with DAS levels is quantified. Secondly, a comparison of different learning methods is outlined. Thirdly, within the overall best performing machine learning model we compare perfomance of the binary and weighted BOW representations of recalls, using only the embedding coming from network centrality. We then provide key results about several models using different combinations of network distances, further enriching the ERT data with featuring coming from network navigation of semantic memory (see Methods). We conclude with the application of the best performing model to the analysis of suicide letters and present the results of psycholinguistic validation of DASentimental estimations.

**Semantic distances reflect patterns of depression, anxiety and stress**

We find that semantic distances, in the network representation of semantic memory, correlate with DAS levels. In other words, the emotional words produced by individuals tend to be closer or further apart to targets like "depression", "anxiety" etc. (see Methods) according to the DAS levels recorded via the psychometric scale. We find a Pearson's correlation coefficient $R$ between depression levels and total semantic distance between recalls and "depression" equal to −0.341 ($N = 200, p < 0.0001$). This means that people affected by higher levels of depression tend to recall and produce emotional words that are semantically closer and thus more related to [38, 37] the concept "depression" in semantic memory. We find analogous patterns for "anxiety" and anxiety levels (−0.218, $N = 200, p = 0.002$) and for "stress" and stress levels (−0.357, $N = 200, p < 0.0001$). These signals provide quantitative evidence that semantic distance from these target concepts can be useful features for predicting DAS levels. For the happy/sad emotional dimension, we find that people with higher depression/anxiety/stress levels tend to produce concepts closer to "sad" ($R < -0.209, N = 200, p < 0.001$). Only people affected by lower depression levels tended to recall items closer to "happy" ($R = 0.162, N = 200, p = 0.02$), whereas no statistically significant correlations were found for anxiety and stress levels. These results indicate that the sad/happy dimensions might be particularly relevant for the estimation of depression levels. At a significance level of 0.05, no other correlations were found. Nonetheless, there might be additional correlations between different features and exploitable by machine learning, so that we will further test the relevance of distance over machine learning regression within the trained models.

The relevance of semantic distances in predicting DAS levels can be visualised also by partitioning the ERT dataset into individuals with higher-than-median (high) or lower-than-median depression (anxiety, stress) levels. Figure 2 (bottom) shows that people with lower levels tend to produce distributions of network distances differing in their medians (Kruskal Wallis test, $N_H + N_L = 200$, $p < 0.001$). Individuals with higher levels of depression (anxiety, stress) tend to recall items more semantically prominent to "depression" ("anxiety", "stress"), further validating the above correlation analysis and the adoption of network distance as features for regression.

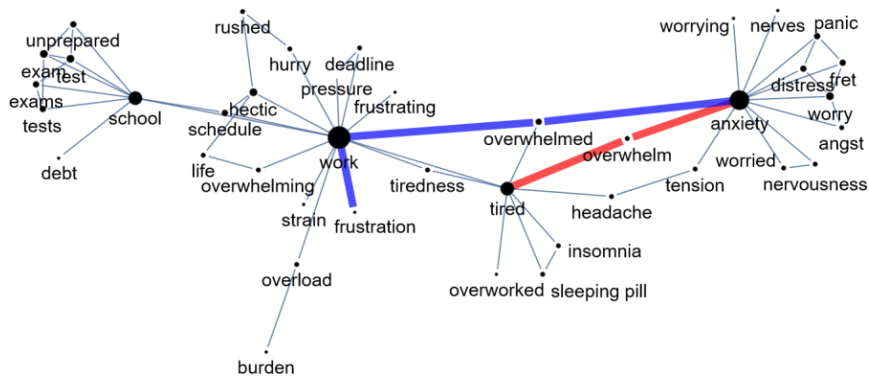

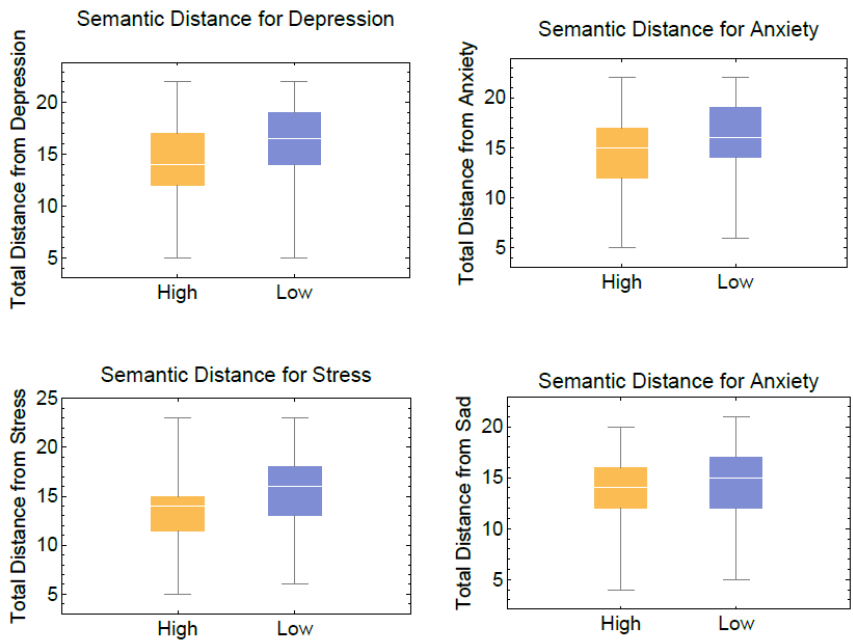

Figure 2: **Top:** Toy representation of semantic memory as a network of free associations. For a recall "tired","frustration",..., the semantic network distances to "anxiety" are highlighted. This visualisation conveys the idea that cognitive networks provide structure to conceptual organisation in the mental lexicon and enable measurements like semantic relatedness in terms of shortest paths/network distance. **Bottom:** Total network distances between recalls and individual concepts (i.e. "anxiety", "depression", "stress" and "sad") between people with high and low levels of DAS.

**Performance of different machine learning algorithms**

As reported in the Methods, we trained ERT data through 3 machine learning algorithms: (i) decision trees, (ii) LSTM recurrent neural networks and (iii) multi-layer perceptron. Independently on using the binary or embedded bag-of-words representations of ERT recall and tuning hyperparameters, neither decision trees nor LSTM networks managed to learn from the dataset. This might be due to the relatively small size of the current sample (200 recalls). It must be also noticed that decision trees try to split the data based on single feature value, whereas in the current case DAS levels might depend on the co-existence of emotional words. For instance, the sequence 'love, broken' might portray love in a painful context, creating a co-dependence of features that would be difficult for decision trees to account for. Instead, the multi-layer perceptron managed to learn relationships from the data and its performance is outlined in the next section in terms of mean-square-error loss and $R^2$ between estimations and validation values of DAS.

**Embedding BOW in semantic memory significantly boosts regression performance**

Table 1 reports the average performance of the multi-layer perceptron regressor over binary and weighted representations of word recalls from the ERT. Notice that in our approach, weights come from the median centralities of words in the network representation of semantic memory enabled by free associations [39, 19, 38] (see Methods).

We find that the binary Bag-Of-Words representation of recalls achieves non-trivial regression results ($R^2$ higher than 0) only for the estimation of depression levels, whereas it fails to do so for both anxiety and stress estimation. Enriching the very same vector representation with the weights coming from cognitive network science drastically boosts performance, with $R^2$ ranging between 0.15 and 0.40 (i.e. $R$ between 0.38 and 0.63). These results indicate that the first items recalled from semantic memory possess more information about the DAS levels of a given individual. This indicates that there is additional structure within the ERT data, that we keep on capitalising by using the weighted representation for our further investigations.

| DAS Constructs | Binary BOW | | Cognitive Weighted BOW | |
| --- | --- | --- | --- | --- |
| | MSE Loss | $R^2$ | MSE Loss | $R^2$ |
| Depression | 30.7 ± 0.1 | 0.19 ± 0.01 | 22.0 ± 0.1 | 0.40 ± 0.02 |
| Anxiety | 16.2 ± 0.1 | 0.03 ± 0.01 | 14.5 ± 0.1 | 0.15 ± 0.02 |
| Stress | 27.6 ± 0.1 | 0.03 ± 0.01 | 19.3 ± 0.1 | 0.26 ± 0.01 |

Table 1: Average losses and $R^2$ estimators for the binary and weighted versions of Bag-Of-Words (BOW) representations of ERT recalls. Weights were fixed according to the median centralities of words in each position of the ERT data (see Methods). Error margins are computed over 10 iterations and indicate standard deviations.

## 1.1 Comparison of model performance based on cognitive network features

Table 2 reports model performance when different network distances are plugged in together with the weighted BOW.

Considering all semantic distances is beneficial in boosting regression results, reducing the MSE loss and enhancing $R^2$ levels, up to 0.49 for depression ($R = 0.7$), 0.20 for anxiety ($R = 0.44$) and 0.27 for stress ($R = 0.52$) levels. Notice that the artificial intelligence trained here, with no additional psychometric information about individuals, correlates as strongly as the ERT metric introduced by Li and colleagues [8]. Hence, we consider DASentimental as working at the state-of-the-art in assessing depression, anxiety and stress levels from emotional recall data, and can proceed using it for text analysis.

Table 2 is important also because it identifies the relevance of different distance-based features in predicting DAS levels. We find that the addition of coverage is beneficial to boosting prediction performance compared to the weighted BOW only, which might be due to non-linear effects that cannot be captured by the previous regression analysis. Adding happy/sad distances to coverage worsens prediction results for "depression" and, in general, in produces a lower boost than adding distances from depression/stress/anxiety to the unweighted BOW representation. Adding all these distances introduces feature correlations that are exploited by the multi-layer perceptron for achieving higher performance.

Table 2 also reports crucial results for exploring how "fear" relates with the estimation of DAS levels. The semantic distances from fear produce a boost in predicting anxiety. This indicates that fear is an important emotion for the prediction of anxiety levels. No boost was recovered for other DAS constructs. However, these distances are correlated with other ones, so that the two models using "fear" and all other concepts provide equivalent performance to the simpler model without fear. For this reason, we selected as the final model of DASentimental the one based on the weighted BOW representation plus coverage/entropy and all other distances except from fear.

## 1.2 Analysis of Suicide Notes

According to the World Health Organization[1], every year more than 700.000 people terminate their lives. Not only does suicide affect the victims, but it also has a trailing effect on their loved ones. People do not commit suicide due to a single reason, but it can be a butterfly effect: starting from minor incidents accumulating over time, leading to increasing mental distress and the appearance of stress, anxiety and/or depression, culminating in the incapability of handling excessive mental pressure and, lastly, triggering the decision of ending one's own life [5].

---

[1] See: https://www.who.int/news-room/fact-sheets/detail/suicide, Last Accessed: 5 October 2021.

| Cognitive-Network Embedded BOW with: | Construct | MSE Loss | $R_2$ |
|---|---|---|---|
| All Conceptual Distances + Cover. + Entr. | Depression | 18.6±0.4 | 0.49±0.01 |
|  | Anxiety | 14.3±0.3 | 0.20±0.02 |
|  | Stress | 19.3±0.3 | 0.27±0.01 |
| Only Distances from Depression/Anxiety/Stress + Cover. + Entr. | Depression | 19.5±0.5 | 0.46±0.01 |
|  | Anxiety | 14.4±0.3 | 0.17±0.02 |
|  | Stress | 19.0±0.4 | 0.28±0.02 |
| Only Distances from Happy/Sad + Cover. + Ent. | Depression | 20.6±0.5 | 0.43±0.01 |
|  | Anxiety | 14.9±0.2 | 0.15±0.01 |
|  | Stress | 19.3±0.4 | 0.27±0.01 |
| Only Cover. + Entr. | Depression | 19.5±0.6 | 0.45±0.01 |
|  | Anxiety | 14.7±0.2 | 0.15±0.01 |
|  | Stress | 19.2±0.4 | 0.27±0.02 |
| Cover. + Entr. + All Distances except from Fear | Depression | 18.5±0.3 | 0.49±0.01 |
|  | Anxiety | 13.9±0.3 | 0.23±0.02 |
|  | Stress | 18.9±0.5 | 0.28±0.01 |
| Distance from Fear only | Anxiety | 14.6±0.2 | 0.16±0.01 |

Table 2: Average losses and $R^2$ estimators for different models employing different features within the same neural network architecture. All the models include weighted representations of words, in addition they feature either: (i) all network distances/conceptual entries (i.e. from "anxiety", "depression", "stress", "sad", "happy" and "fear", together with the total emotional coverage), or (ii) only distances from "depression", "stress" and "anxiety" with coverage and graph distance entropy, or (iii) only distances from "happy" and "sad" with the coverage and entropy, or (iv) only coverage and entropy, or (v) all other distances, coverage and entropy but without distances from "fear", or (vi) only distances from "fear". Error margins are computed over 10 iterations and indicate standard deviations.

Though the count of suicides is high, the number of people leaving suicide notes is just a fraction of this. Suicide notes are the vital piece of information, that can give insight on the vulnerable mindset of the individual committing suicide [28, 29]. These notes are written by the individuals who have reached the limit of emotional distress. These are the first-hand evidence of the mindset of emotionally distraught individuals, so that analysing these notes can provide important insights over the mental distress of their authors.

To gather such insights, we applied DASentimental, i.e. its best performing version with weighted BOW and semantic distances, over the corpus of genuine suicide notes curated by Schoene and Deathlefs [28] and investigated in other recent studies [29, 24]. Notice that this application constitutes the second part of our semi-supervised approach to text analysis, where DASentimental predicts DAS levels of nonannotated text from its semantically enriched sequences of emotional words (see Methods).

Results are reported in Figure 3. We registered strong positive correlations between estimated DAS levels (Pearson's coefficients, $R_{DA} = 0.35$, $p < 0.0001$; $R_{DS} = 0.50$, $p < 0.0001$; $R_{AS} = 0.59$, $p < 0.0001$). These indicate that suicide notes tended to feature analogous levels of distress coming from anxiety, depression and stress, although with different intensities and frequencies, as evident from the qualitative analysis of distributions in Fig. 3 (bottom).

Using valence and arousal of words expressed in suicide letters, we performed an additional validation of the results of DAS through the circumplex model of affect [11] (see Methods). By partitioning the notes according to high/low levels of anxiety (depression, stress), we compared the valence (and arousal) of all words mentioned in suicide letters from each partition. At a significance level of 0.05, suicide notes marked by DASentimental with higher depression levels were found to contain a lower median valence than notes marked with lower levels of depression (Kruskal Wallis test, $KS = 6.889$, $p = 0.009$). Analogously, suicide notes marked by the AI with higher anxiety are found to contain a higher median arousal than notes marked with lower anxiety (Kruskal Wallis test, $KS = 3.2014$, $p = 0.007$). No differences for stress were found.

Letters with lower depression levels were found to contain more positive jargon, including mentions of loved ones and "relief" for ending the pain and starting a new chapter. Some notes even included emotionless instructions about relatives and assets needed to be taken care of. Letters with higher depression levels mentioned more frequently jargon relative to 'pain' and 'boredom' and this imbalance in frequency is captured by the above difference in median valence. Since in the circumplex model, depression lives in a space with more negative valence than neutral/emotionless language, the above statistically significant difference in valence indicates that DASentimental is able to identify the negative dimension associated to depression.

An analogous pattern was found for anxiety, with letters tagged as "high anxiety" by the AI featuring more anxious jargon relative to pain and suffering. Since in the circumplex model, anxiety lives in a space with higher arousal and alarmness than neutral/emotionless language, the above difference in median arousal between high- and low-anxiety letters indicates that DASentimental is able to identify the alarming and arousal-inducing dimension associated to anxiety.

The absence of differences for stress might indicate either that the AI is not powerful enough to detect differences in stress, underlining the need for future research and larger datasets. Nonetheless, the signals of enhanced negativity and alarmness detected by DASentimental lay the foundation for an interesting starting point for detecting stress, anxiety and depression in texts via emotional recall data.

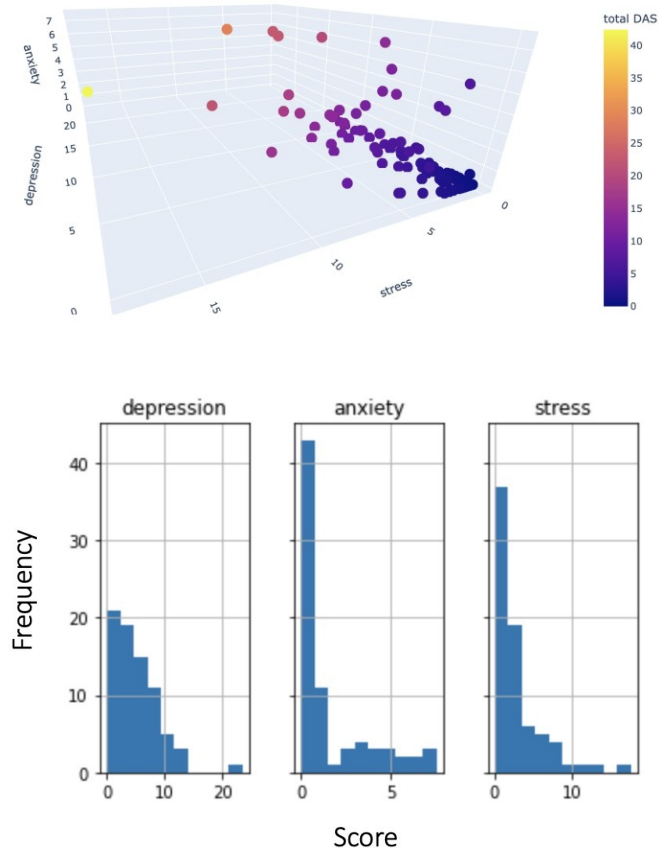

Figure 3: Top: 3D visualisation of depression, anxiety and stress of suicide notes as estimated by DASentimental. Bottom: Histograms of DAS levels per pathological construct.

## Discussion

In this study, we trained a neural network (DASentimental) to predict depression, anxiety and stress starting from sequences of emotional words embedded in a cognitive network representation of semantic memory. We found that DASentimental achieves cross-validated predictions for depression (R = 0.7), anxiety (R = 0.44), and stress (R = 0.52) in line with previous approaches using additional valence data [8]. This state-of-art performance suggests that even without explicitly encoding valence ratings for each word, the artificial intelligence is able to achieve good explanatory power, rising from the cognitive embedding of recalled concepts, i.e. concepts being at shorter/longer network distance from key ideas like "depression", "sad", "happy", etc. that are commonly used to describe emotional distress [4, 7].

Our findings suggest the importance of considering network distances between concepts in semantic memory for investigating emotional distress, providing further support to other studies showing how network distances and connectivity can predict other cognitive phenomena like creativity levels [16, 18, 20], semantic distance [38, 37] and word production in clinical populations [36].

We noticed a significant boost in performance (+210% in $R^2$ on average) when embedding Bag-OfWords representations of recall lists (see Methods) in a cognitive network of free associations [33]. A boost of almost ten times was observed for predicting stress and anxiety, which are considerably complex distress constructs [14, 40]. Our results underline the need to tie together artificial intelligence/textmining [43] and cognitive network science [15] for achieving cutting-edge predictors in next-generation cognitive computing.

We applied DAsentimental to a collection of suicide notes as a case study. Most suicide notes in the corpus [28] were relative to low levels of depression, anxiety, and stress. This suggests that despite the decision to terminate one's own life, the writers of suicide notes tried to avoid overwhelming their last messages with negative emotions, compatible with previous studies [24]. One observation coming from reading closely suicide notes is that many writers expressed their love and gratitude to their significant others, and used euphemisms when referencing the act of suicide (e.g. I can't carry this anymore). Therefore, although readers would find a typical suicide note filled with sorrow, that perception is built on the contextual knowledge that the writer eventually killed him/herself. A key limitation of DASentimental is that it cannot account for linguistic pragmatics, i.e. how context shifts and forges meaning and perceptions in language [43]. Furthermore, DASentimental cannot capture how the writers actually felt before, during or after writing those last letters. Instead, we argue that DASentimental quantifies those emotions as explicitly expressed by the authors, since it is trained on ERT data which includes expressions of emotions without context. Future research might better detect contextual knowledge through natural language processing, which has been successfully used to detect the risk of psychosis in clinical populations coming from contextual features like medical reports [52] or speech organisation [27]. Alternatively, community detection in feature-rich networks could inform over the different meanings/contextual interpretations provided to concepts in cognitive networks, as showcases by Citraro and Rossetti [53] for the different meanings of "star". Last but not least, contextual features might be detected through meso-scale network metrics like entanglement, which was recently shown to efficiently identify nodes critical for information diffusion in a variety of techno-social networks [54].

Notice that DASentimental uses cognitive network distances [45] to target words, a feature that replaces the subjective valence ratings adopted in the original ERT study [8], ratings not being available in texts. Despite this difference, DASentimental obtains analogous performance with the work of Li and colleagues [8]. This has two implications: (i) a better model might use distances and valences in the future when focusing only on fluency tasks, and (ii) for text analysis and even cognitive social media mining [55], the machine learning pipeline of DASentimental could be used to detect *any kind of target emotion* (e.g., 'surprise' or 'love').

As a future research direction, one promising application of DASentimental is investigating the cultural evolution of emotions. Emotions and their expressions are shaped by culture and learned in social contexts [56, 57] and media movements [49]. What people can feel and express depends on their surrounding social norms. Previous studies have shown large historical corpora can be used to make quantitative inferences on the rise and fall of national happiness [57]. Similarly, DASentimental could be applied to track the change of explicit expression of depression, anxiety, and stress over history, quantified through emotions of "modern" individuals. This would highlight changes in norms towards emotional expression and historical events such as "pandemic", complementing other recent cognitive network science [30, 58, 59, 9, 60] and sentiment/emotional profiling [51, 61, 55, 62] approaches by bringing on the table a quantitative, automatic quantification of anxiety, stress and depression in texts.

## References

[1] Peter F Lovibond and Sydney H Lovibond. The structure of negative emotional states: Comparison of the depression anxiety stress scales (dass) with the beck depression and anxiety inventories. Behaviour research and therapy, 33(3):335–343, 1995.

[2] James A Russell and Lisa Feldman Barrett. Core affect, prototypical emotional episodes, and other things called emotion: dissecting the elephant. Journal of personality and social psychology, 76(5):805, 1999.

[3] Ciar´an O'Driscoll, Joshua EJ Buckman, Eiko I Fried, Rob Saunders, Zachary D Cohen, Gareth Ambler, Robert J DeRubeis, Simon Gilbody, Steven D Hollon, Tony Kendrick, et al. The importance of transdiagnostic symptom level assessment to understanding prognosis for depressed adults: analysis of data from six randomised control trials. BMC medicine, 19(1):1–14, 2021.

[4] Ahmet Akin and Bayram C¸etın. The depression anxiety and stress scale (dass): The study of validity and reliability. Educational Sciences: Theory & Practice, 7(1), 2007.

[5] Ismael Conejero, Emilie Oli´e, Raffaella Calati, D´eborah Ducasse, and Philippe Courtet. Psychological pain, depression, and suicide: recent evidences and future directions. Current psychiatry reports, 20(5):1–9, 2018.

[6] Rany Abend, Mira A Bajaj, Daniel DL Coppersmith, Katharina Kircanski, Simone P Haller, Elise M Cardinale, Giovanni A Salum, Reinout W Wiers, Elske Salemink, Jeremy W Pettit, et al. A computational network perspective on pediatric anxiety symptoms. Psychological medicine, 51(10):1752–1762, 2021.

[7] Lisa Feldman Barrett. Valence is a basic building block of emotional life. Journal of Research in Personality, 40(1):35–55, 2006.

[8] Ying Li, Annasya Masitah, and Thomas T Hills. The emotional recall task: Juxtaposing recall and recognition-based affect scales. Journal of Experimental Psychology: Learning, Memory, and Cognition, 2020.

[9] Maria Montefinese, Ettore Ambrosini, and Alessandro Angrilli. Online search trends and wordrelated emotional response during covid-19 lockdown in italy: a cross-sectional online study. PeerJ, 9:e11858, 2021.

[10] Saif Mohammad. Obtaining reliable human ratings of valence, arousal, and dominance for 20,000 english words. In Proceedings of the 56th Annual Meeting of the Association for Computational Linguistics (Volume 1: Long Papers), pages 174–184, 2018.

[11] Jonathan Posner, James A Russell, and Bradley S Peterson. The circumplex model of affect: An integrative approach to affective neuroscience, cognitive development, and psychopathology. Development and psychopathology, 17(3):715–734, 2005.

[12] Auke Tellegen. Structures of mood and personality and their relevance to assessing anxiety, with an emphasis on self-report. 1985.

[13] David Watson, Lee Anna Clark, and Auke Tellegen. Development and validation of brief measures of positive and negative affect: the panas scales. Journal of personality and social psychology, 54(6):1063, 1988.

[14] Michele M Tugade, Barbara L Fredrickson, and Lisa Feldman Barrett. Psychological resilience and positive emotional granularity: Examining the benefits of positive emotions on coping and health. Journal of personality, 72(6):1161–1190, 2004.

[15] Yoed N Kenett and Miriam Faust. Clinical cognitive networks: A graph theory approach. In Network science in cognitive psychology, pages 136–165. Routledge, 2019.

[16] Roger E Beaty, Daniel C Zeitlen, Brendan S Baker, and Yoed N Kenett. Forward flow and creative thought: Assessing associative cognition and its role in divergent thinking. Thinking Skills and Creativity, page 100859, 2021.

[17] Cynthia SQ Siew, Dirk U Wulff, Nicole M Beckage, and Yoed N Kenett. Cognitive network science: A review of research on cognition through the lens of network representations, processes, and dynamics. Complexity, 2019.

[18] Yoed N Kenett, Orr Levy, Dror Y Kenett, H Eugene Stanley, Miriam Faust, and Shlomo Havlin. Flexibility of thought in high creative individuals represented by percolation analysis. Proceedings of the National Academy of Sciences, 115(5):867–872, 2018.

[19] Abhilasha A Kumar. Semantic memory: A review of methods, models, and current challenges. Psychonomic Bulletin & Review, 28(1):40–80, 2021.

[20] Massimo Stella and Yoed N Kenett. Viability in multiplex lexical networks and machine learning characterizes human creativity. Big Data and Cognitive Computing, 3(3):45, 2019.

[21] Thomas T Hills, Michael N Jones, and Peter M Todd. Optimal foraging in semantic memory. Psychological review, 119(2):431, 2012.

[22] Hudson F Golino and Sacha Epskamp. Exploratory graph analysis: A new approach for estimating the number of dimensions in psychological research. PloS one, 12(6):e0174035, 2017.

[23] Jorge AV Tohalino, Laura VC Quispe, and Diego R Amancio. Analyzing the relationship between text features and grants productivity. Scientometrics, 126(5):4255–4275, 2021.

[24] Andreia Sofia Teixeira, Szymon Talaga, Trevor James Swanson, and Massimo Stella. Revealing semantic and emotional structure of suicide notes with cognitive network science. arXiv preprint arXiv:2007.12053, 2020.

[25] Jeffrey C Zemla, Kesong Cao, Kimberly D Mueller, and Joseph L Austerweil. Snafu: The semantic network and fluency utility. Behavior research methods, 52(4):1681–1699, 2020.

[26] Sarah E Morgan, Kelly Diederen, Petra E Vertes, Samantha HY Ip, Bo Wang, Bethany Thompson, Arsime Demjaha, Andrea De Micheli, Dominic Oliver, Maria Liakata, et al. Assessing psychosis risk using quantitative markers of disorganised speech. medRxiv, 2021.

[27] Sarah E. Morgan, Kelly Diederen, Petra E. V´ertes, Samantha Ip, Bo Wang, Bethany Thompson, Arsime Demjaha, Andrea De Micheli, Dominic Oliver, Maria Liakata, Paolo Fusar-Poli, Tom Spencer, and Philip McGuire. Natural language processing markers in first episode psychosis and people at clinical high-risk. Translational Psychiatry, September 2021.

[28] Annika Marie Schoene and Nina Dethlefs. Automatic identification of suicide notes from linguistic and sentiment features. In Proceedings of the 10th SIGHUM Workshop on Language Technology for Cultural Heritage, Social Sciences, and Humanities, pages 128–133, 2016.

[29] Annika Marie Schoene, Alexander Turner, Geeth Ranmal De Mel, and Nina Dethlefs. Hierarchical multiscale recurrent neural networks for detecting suicide notes. IEEE Transactions on Affective Computing, 2021.

[30] Massimo Stella, Valerio Restocchi, and Simon De Deyne. # lockdown: Network-enhanced emotional profiling in the time of covid-19. Big Data and Cognitive Computing, 4(2):14, 2020.

[31] Thorsten Pachur, Ralph Hertwig, and Florian Steinmann. How do people judge risks: availability heuristic, affect heuristic, or both? Journal of Experimental Psychology: Applied, 18(3):314, 2012.

[32] Amos Tversky and Daniel Kahneman. Availability: A heuristic for judging frequency and probability. Cognitive psychology, 5(2):207–232, 1973.

[33] Simon De Deyne, Danielle J Navarro, Amy Perfors, Marc Brysbaert, and Gert Storms. The "small world of words" english word association norms for over 12,000 cue words. Behavior research methods, 51(3):987–1006, 2019.

[34] Thomas T Hills, Mounir Maouene, Josita Maouene, Adam Sheya, and Linda Smith. Longitudinal analysis of early semantic networks: Preferential attachment or preferential acquisition? Psychological science, 20(6):729–739, 2009.

[35] Massimo Stella, Nicole M Beckage, and Markus Brede. Multiplex lexical networks reveal patterns in early word acquisition in children. Scientific reports, 7(1):1–10, 2017.

[36] Nichol Castro, Massimo Stella, and Cynthia SQ Siew. Quantifying the interplay of semantics and phonology during failures of word retrieval by people with aphasia using a multiplex lexical network. Cognitive Science, 44(9):e12881, 2020.

[37] Abhilasha A Kumar, David A Balota, and Mark Steyvers. Distant connectivity and multiple-step priming in large-scale semantic networks. Journal of Experimental Psychology: Learning, Memory, and Cognition, 46(12):2261, 2020.

[38] Yoed N Kenett, Effi Levi, David Anaki, and Miriam Faust. The semantic distance task: Quantifying semantic distance with semantic network path length. Journal of Experimental Psychology: Learning, Memory, and Cognition, 43(9):1470, 2017.

[39] Simon De Deyne, Daniel J Navarro, and Gert Storms. Better explanations of lexical and semantic cognition using networks derived from continued rather than single-word associations. Behavior research methods, 45(2):480–498, 2013.

[40] Amy M Smith, Gregory I Hughes, F Caroline Davis, and Ayanna K Thomas. Acute stress enhances general-knowledge semantic memory. Hormones and behavior, 109:38–43, 2019.

[41] Yoed Kenett, Brendan Baker, Thomas Hills, Yuval Hart, and Roger Beaty. Creative foraging: Examining relations between foraging styles, semantic memory structure, and creative thinking. In Proceedings of the Annual Meeting of the Cognitive Science Society, volume 43, 2021.

[42] Fabian Pedregosa, Ga¨el Varoquaux, Alexandre Gramfort, Vincent Michel, Bertrand Thirion, Olivier Grisel, Mathieu Blondel, Peter Prettenhofer, Ron Weiss, Vincent Dubourg, et al. Scikit-learn: Machine learning in python. the Journal of machine Learning research, 12:2825–2830, 2011.

[43] Hossein Hassani, Christina Beneki, Stephan Unger, Maedeh Taj Mazinani, and Mohammad Reza Yeganegi. Text mining in big data analytics. Big Data and Cognitive Computing, 4(1):1, 2020.

[44] Thomas T Hills and Yoed N Kenett. Networks of the mind: How can network science elucidate our understanding of cognition? Topics in Cognitive Science.

[45] Massimo Stella and Manlio De Domenico. Distance entropy cartography characterises centrality in complex networks. Entropy, 20(4):268, 2018.

[46] Ethem Alpaydin. Introduction to machine learning. MIT press, 2020.

[47] Matt W Gardner and SR Dorling. Artificial neural networks (the multilayer perceptron)—a review of applications in the atmospheric sciences. Atmospheric environment, 32(14-15):2627–2636, 1998.

[48] Hendrik Vankrunkelsven, Steven Verheyen, Simon De Deyne, and Gert Storms. Predicting lexical norms using a word association corpus. In Proceedings of the 37th annual conference of the cognitive science society, pages 2463–2468. Cognitive Science Society; Austin, TX, 2015.

[49] Diego Raphael Amancio, Osvaldo N Oliveira Jr, and Luciano da Fontoura Costa. Identification of literary movements using complex networks to represent texts. New Journal of Physics, 14(4):043029, 2012.

[50] Massimo Stella. Text-mining forma mentis networks reconstruct public perception of the stem gender gap in social media. PeerJ Computer Science, 6:e295, 2020.

[51] Saif M Mohammad. Sentiment analysis: Automatically detecting valence, emotions, and other affectual states from text. In Emotion Measurement, pages 323–379. Elsevier, 2021.

[52] Jessica Irving, Rashmi Patel, Dominic Oliver, Craig Colling, Megan Pritchard, Matthew Broadbent, Helen Baldwin, Daniel Stahl, Robert Stewart, and Paolo Fusar-Poli. Using natural language processing on electronic health records to enhance detection and prediction of psychosis risk. Schizophrenia bulletin, 47(2):405–414, 2021.

[53] Salvatore Citraro and Giulio Rossetti. Identifying and exploiting homogeneous communities in labeled networks. Applied Network Science, 5(1):1–20, 2020.

[54] Arsham Ghavasieh, Massimo Stella, Jacob Biamonte, and Manlio De Domenico. Unraveling the effects of multiscale network entanglement on empirical systems. Communications Physics, 4(1):1– 10, 2021.

[55] Massimo Stella. Cognitive network science for understanding online social cognitions: A brief review. Topics in Cognitive Science, page 12551.

[56] Virginia Morini, Laura Pollacci, and Giulio Rossetti. Toward a standard approach for echo chamber detection: Reddit case study. Applied Sciences, 11(12):5390, 2021.

[57] Thomas T Hills, Eugenio Proto, Daniel Sgroi, and Chanuki I Seresinhe. Historical analysis of national subjective wellbeing using millions of digitized books. Nature human behaviour, 3(12):1271–1275, 2015.

[58] Felix M Simon and Chico Q Camargo. Autopsy of a metaphor: The origins, use and blind spots of the 'infodemic'. New media & Society, page 14614448211031908.

[59] Ying Li, Shenghua Luan, Yugang Li, and Ralph Hertwig. Changing emotions in the covid-19 pandemic: A four-wave longitudinal study in the united states and china. Social Science & Medicine, 285:114222, 2021.

[60] Matteo Cinelli, Walter Quattrociocchi, Alessandro Galeazzi, Carlo Michele Valensise, Emanuele Brugnoli, Ana Lucia Schmidt, Paola Zola, Fabiana Zollo, and Antonio Scala. The covid-19 social media infodemic. Scientific Reports, 10(1):1–10, 2020.

[61] Alfonso Semeraro, Salvatore Vilella, and Giancarlo Ruffo. Pyplutchik: visualising and comparing emotion-annotated corpora. Plos One, page 0256503, 2021.

[62] Tommaso Radicioni, Tiziano Squartini, Elena Pavan, and Fabio Saracco. Networked partisanship and framing: a socio-semantic network analysis of the italian debate on migration. arXiv preprint arXiv:2103.04653, 2021.

## Supplementary Information

This Supplementary Information provides two examples of suicide notes selected uniformly at random from the high/low partitions discussed in the main text.

1. Suicide note reported with **high** depression anxiety and stress.

   Content: 'Dearest Elinor. This is to say goodbye. I have not told you because I did not want to worry you but I have been feeling bad for 25 years with my heart. I knew that if I went to a doctor I would lose my job. I think this is best for al concerened. I am in the car in the garage. Call the police but please don't come out there. I love you very much darling. Goodbye William'

   Assignments: Table 3

   DAS Scores: Depression: 4.04 Anxiety: 10.2 Stress: 26

   | Word in Suicide Note | Assigned word from ERT dataset | Similarity |
   |---|---|---|
   | goodbye | sad | 0.560979152 |
   | worry | worry | 1 |
   | bad | horrible | 0.761918145 |
   | heart | love | 0.526728102 |
   | knew | understand | 0.838536982 |
   | doctor | patient | 0.669662141 |
   | lose | lose | 1 |
   | job | good | 0.592258104 |
   | think | wonder | 0.827612808 |
   | best | good | 0.708921861 |
   | car | drive | 0.505276574 |
   | police | suspicious | 0.531217872 |
   | love | love | 1 |
   | darling | love | 0.524814114 |

   Table 3: Word assignment for each non-stop word in suicide note 1

   | Word in Suicide Note | Assigned word from ERT dataset | Similarity |
   |---|---|---|
   | check | please | 0.600356847 |
   | withdraw | withdraw | 1 |
   | money | spend | 0.644925672 |

| Thank | grateful | 0.759848329 |
| Pay | spend | 0.660345671 |
| business | successful | 0.542733439 |
| found negation 'not' | | |
| neg sold - buy | cheap | 0.746127929 |
| houses | family | 0.560697799 |
| time | wait | 0.662202806 |
| house | family | 0.560697799 |

Table 4: Word Assignment for each non-stop word in Suicide note 2

2. Suicide note reported with **low** depression anxiety and stress scores.

    Content: 'Jane give all of my possessions to Elinor and dont want Lizzie to attend my funeral. William [] Elinor Please take this check and withdraw all the money from my account Thank you William Please Pay Christopher at ( business ) $ 2000 ( tel. BA 00000 ) The vW License[] should be retrieved from ( auto shop ) across the street and given to Jane when she turns 30 and is not to be sold My share of the houses should go to Jane and she is to retain full possession until such time the share of this house that belongs to me should go to alcoholics Anonamous'

    Assignments: Table 4

    DAS Scores: Depression: 0 Anxiety: 0 Stress: 0